\documentclass{iau}
\usepackage{graphicx}
\usepackage{amsmath,amsfonts,amssymb}

\newcommand{\avg}[1]{\left\langle #1 \right\rangle}

\def\BV{\hbox{$\avg{\rm B_{\rm V}}$}}

\def\vsini{\hbox{$v\sin i$}}

\def\vec#1{\ensuremath{\mathchoice{\mbox{\boldmath$\displaystyle#1$}}
{\mbox{\boldmath$\textstyle#1$}}
{\mbox{\boldmath$\scriptstyle#1$}}
{\mbox{\boldmath$\scriptscriptstyle#1$}}}}

\title[What controls the large-scale magnetic fields of M dwarfs?]
{What controls the large-scale magnetic fields of M dwarfs?}

\author[T. Gastine, J. Morin, L. Duarte, A. Reiners, U. Christensen and J. 
Wicht] {T. Gastine$^1$, J. Morin$^{2,3}$,
L. Duarte$^1$, A. Reiners$^2$, U. Christensen$^1$ and J. Wicht$^1$}

\affiliation{$^1$Max Planck Institut f\"ur Sonnensystemforschung,
\\ Max Planck Stra{\ss}e 2,
37191 Katlenburg-Lindau, Germany		\\ email: {\tt gastine@mps.mpg.de} 
\\[\affilskip]
$^2$Institut f\"ur Astrophysik, Georg-August-Universit\"at G\"ottingen, \\ 
Friedrich-Hund Platz, 37077 G\"ottingen, Germany \\[\affilskip]
$^{3}$  LUPM, Universit\'e de Montpellier and CNRS, \\
Place E. Bataillon, 34090 Montpellier, France}

\pubyear{2013}
\volume{302}  
\pagerange{119--123}
\jname{Magnetic fields throughout stellar evolution}
\editors{M. Jardine, P. Petit \& H. Spruit, eds.}
\begin{document}

\maketitle

\begin{abstract}
Observations of active M dwarfs show a broad variety of large-scale magnetic 
fields encompassing dipole-dominated and multipolar geometries. We detail the 
analogy between some anelastic dynamo simulations and spectropolarimetric 
observations of 23 M stars. In numerical models, the relative contribution of 
inertia and Coriolis force --estimated by the so-called local Rossby number-- is 
known to have a strong impact on the magnetic field geometry. We discuss the 
relevance of this parameter in setting the large-scale magnetic field of M 
dwarfs.

\end{abstract}

\firstsection 
\section{Introduction}

The magnetic fields of planets and rapidly-rotating stars are maintained by
convection-driven dynamos operating in their interiors. Scaling laws recently
derived from geodynamo-like models successfully predict the magnetic field 
strength of a wide range of astrophysical objects from Earth and Jupiter to some
rapidly-rotating stars \cite[(e.g. Christensen \& Aubert 2006; Christensen 
et al. 2009; Yadav et al. 
2013a,b)]{Christensen06,Christensen10,Yadav13a,Yadav13b}. This emphasises the 
similarities between the dynamo
mechanisms at work in planets and active M dwarfs.

Spectropolarimetric observations of rapidly-rotating M stars show a broad
variety of large-scale magnetic fields encompassing dipole-dominated and
multipolar geometries \cite[(Donati et al. 2008; Morin et al. 
2008a,b,2010)]{Donati08,Morin08a,midM,lateM}.
Combining global-scale numerical dynamo models and
observational results, we want to better understand the similarities of dynamos
in planets and low-mass stars. To study the physical mechanisms that control
the magnetic field morphology in these objects, we have explored the influence
of rotation rate, convective vigor and density stratification on the magnetic 
field properties in anelastic dynamo models \cite[(Gastine et al. 
2012,2013)]{Gastine12a,Gastine13}.

In such models, the relative importance of inertia and Coriolis force in the 
force balance --quantified by the local Rossby number $Ro_l$-- is thought to 
have a strong impact on the magnetic field geometry \cite[(Christensen \& 
Aubert 2006)]{Christensen06}. A sharp transition  between dipole-dominated 
and multipolar dynamos is indeed observed at $Ro_l \simeq 0.1$. However,
\cite{Simitev09} find that both dipolar and multipolar magnetic fields are two 
possible solutions at the same parameter regime, depending on the initial 
condition of the system. As shown by \cite{Schrinner12}, this dynamo 
bistability challenges the $Ro_l$ criterion as the multipolar dynamo branch can 
extend well below the threshold value $Ro_l\simeq 0.1$.

Here we discuss the analogy between the anelastic dynamo models by 
\cite{Gastine12a} and the spectropolarimetric observations of 23 M stars.
The reader is referred to \cite[(Gastine et al. 2013)]{Gastine13} for a more 
comprehensive description of the results.

\section{Dynamo models and  spectropolarimetric observations}

We consider MHD simulations of a conducting anelastic fluid in spherical shells
rotating at a constant rotation rate $\Omega$. A fixed entropy contrast 
$\Delta s$ between the inner and the outer boundary drives the convective 
motions. Our numerical models are computed using the anelastic spectral code 
MagIC \cite[(Wicht 2002, Gastine \& Wicht 2012)]{Wicht02,Gastine12} that has 
been validated against several hydrodynamical and dynamo benchmarks \cite[(Jones 
et al. 2011)]{Jones11}. The governing MHD equations are non-dimensionalised 
using the shell thickness $d=r_o-r_i$ as the reference lengthscale and 
$\Omega^{-1}$ as the time unit.

The solution of a numerical model is then characterised by several diagnostic 
parameters. The rms flow velocity is given by the Rossby number 
$Ro=u_{\text{rms}}/\Omega d$, while the magnetic field strength is 
measured by the Elsasser number $\Lambda=B_{\text{rms}}^2 
/\rho\mu\lambda\Omega$, where $\rho$ is the density, and $\mu$ and $\lambda$ 
are the magnetic permeability and diffusivity. The typical flow lengthscale 
$l$ is defined as $l = \pi d/\bar{\ell}_u$, where 
$\bar{\ell}_u$ is the mean spherical harmonic degree obtained from the kinetic 
energy spectrum \cite[(Christensen \& Aubert 2006; Schrinner et al. 
2012)]{Christensen06,Schrinner12}. Following \cite{Christensen06}, a 
\emph{local Rossby number} $Ro_l=  u_{\text{rms}}/\Omega l $,
can then be used to evaluate the impact of inertia on the magnetic field 
geometry. Finally, the geometry of the surface magnetic field is quantified by 
its dipolarity 
$f_\text{dip}=\vec{B}_{\ell=1,m=0}^{2}(r=r_o)/\sum_{\ell,m}^{\ell_{\text{max}}
}\vec{B}_{\ell, m } ^ { 2 } (r=r_o)$, the ratio of the magnetic energy 
of the dipole to the magnetic energy contained in spherical 
harmonic degrees up to $\ell_{\text{max}}=11$.

We compare these dynamo models with spectropolarimetric observations of 23 
active M dwarfs with rotation period ranging from 0.4 to 19 days. The data 
reduction and analysis is detailed by \cite{Donati06} and \cite[Morin et 
al. (2008a,b,2010)]{Morin08a,midM,lateM}. We derive observation-based 
quantities aimed to reflect the diagnostic parameters employed in the numerical 
models. The \emph{empirical Rossby number} $Ro_\text{emp}= P_\text{rot}/\tau_c$ 
is our best available proxy for $Ro_l$, where $\tau_c$ is the turnover 
timescale of convection based on the rotation-activity relation \cite[(Kiraga \& 
Stepien 2007)]{Kiraga07}.  We define an Elsasser number based on the
averaged unsigned large-scale magnetic field  \BV\ that roughly characterises
the ratio between Lorentz and Coriolis forces. We also consider the fraction of 
the magnetic energy that is recovered in the axial dipole mode in 
Zeeman-Doppler imaging maps \cite[(ZDI, Semel 1989)]{Semel89}. The spatial 
resolution of such maps mostly depends on the projected rotational 
velocity \vsini. The actual degree and order $\ell_\text{max}$ up to which the 
reconstruction can be performed ranges from 4 to 10. We directly 
compare this quantity to the dipolarity employed in numerical models and term 
them both $f_\text{dip}$ in Figs.~\ref{fig:dipl11}-\ref{fig:dipM}.

\section{Results and discussion}

\begin{figure}[htbp]
\centering
 \includegraphics[width=8.5cm]{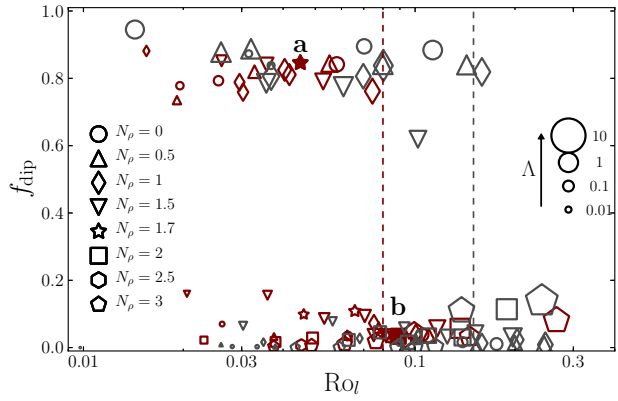} 
 \caption{$f_\text{dip}$ plotted against $Ro_l$ in the anelastic 
dynamo models computed by \cite{Gastine12a}. Red (grey) symbols correspond to 
numerical simulations in thick (thin) shells ($r_i/r_o=0.2$ and $r_i/r_o=0.6$). 
The symbol sizes scale with the amplitude of the surface field, 
given in units of the square-root of the Elsasser number. The two vertical 
lines mark the possible upper-limits of the dipole-dominated dynamos. The two 
filled symbols are further discussed in Fig.~\ref{fig:example}.}
   \label{fig:dipl11}
\end{figure}

\begin{figure}[htbp]
\centering
 \includegraphics[width=8.5cm]{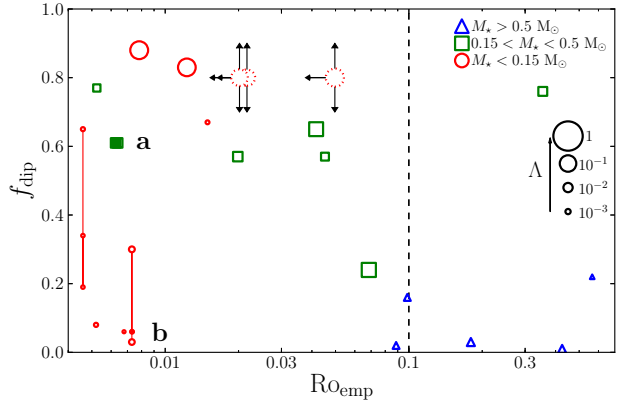} 
 \caption{$f_\text{dip}$ plotted against $Ro_\text{emp}$. The symbol 
sizes scale with the square root of the Elsasser number based on the 
large-scale magnetic field derived from spectropolarimetric observations. The 
vertical dashed line marks the possible upper bound of the dipolar regime. For 
the two stars with the largest temporal variation, individual epochs are 
connected by a vertical red line. Dotted red circles with errorbars correspond 
to some stars from \cite{lateM} for which a definite ZDI reconstruction was not 
possible.}
   \label{fig:dipM}
\end{figure}

\begin{figure}[htbp]
 \centering
 \includegraphics[width=11cm]{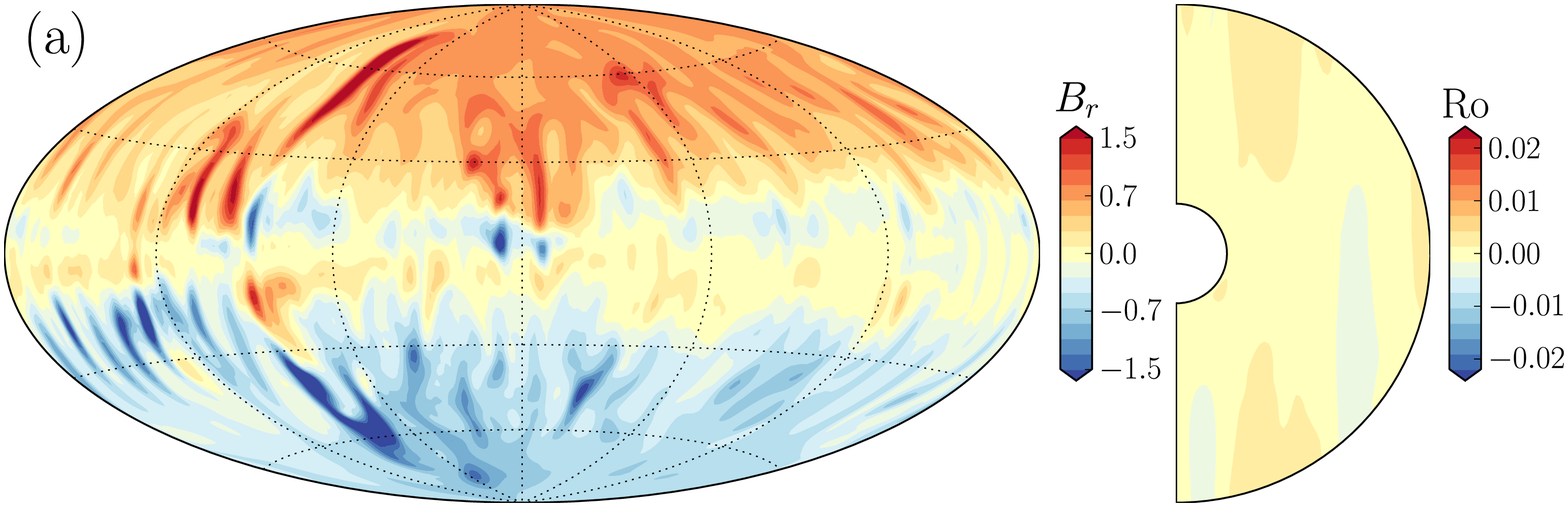}
 \includegraphics[width=11cm]{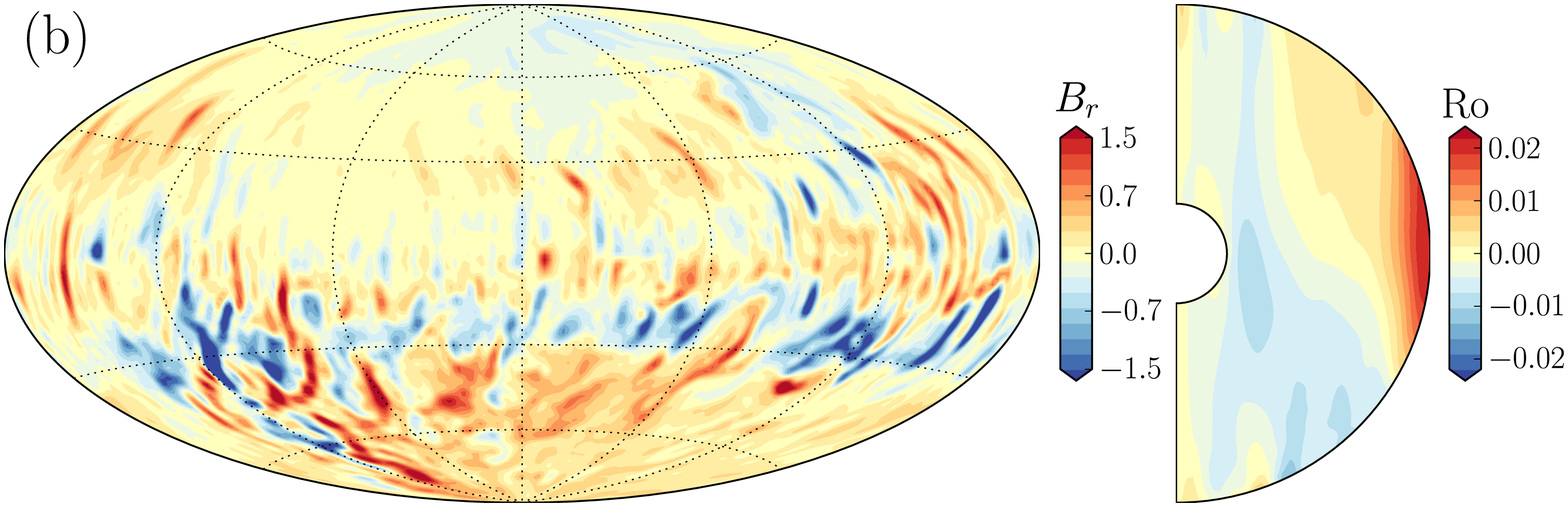}
 \caption{Snapshot of the radial component of the surface magnetic field  
and the axisymmetric zonal flow $\bar{u}_\phi$ for a dipolar dynamo model 
(\textbf{a}), and a multipolar case (\textbf{b}). Magnetic field are given in 
units of the square root of the Elsasser number and velocities in units of the 
Rossby number.}
 \label{fig:example}
\end{figure}

Figure~\ref{fig:dipl11} shows $f_\text{dip}$ versus $Ro_l$ in the
numerical models, while Fig.~\ref{fig:dipM} displays the relative dipole 
strength of M stars against $Ro_\text{emp}$ derived from spectropolarimetric 
observations. The numerical models cluster in two distinct dynamo branches: the 
upper branch corresponds to the dipole-dominated regime ($f_\text{dip} > 0.6$), 
while the lower branch contains the multipolar dynamos ($f_\text{dip} < 0.2$). 
Fig.~\ref{fig:example} shows two selected cases of these two kinds of dynamo 
action. The dipolar branch is limited by a maximum $Ro_l \simeq 0.1$, 
beyond which all the models become multipolar. In
contrast to earlier Boussinesq studies \cite[(e.g. Christensen \& Aubert 
2006)]{Christensen06}, the multipolar branch also extends well below 
$Ro_l \simeq 0.1$, where both dipolar and multipolar solutions are 
stable \cite[(see Schrinner et al. 2012)]{Schrinner12}. Bistability of the 
magnetic field is in fact quite common in the parameter range explored here, 
meaning that both dipole-dominated and multipolar fields are two possible 
stable configurations at the same set of parameters \cite[(Simitev 
\& Busse 2009)]{Simitev09}. The multipolar branch at low $Ro_l$ is partly 
composed by the anelastic models with $\rho_\text{bot}/\rho_\text{top} > 7$ 
\cite[(Gastine et al. 2012)]{Gastine12} and partly by the multipolar attractors 
of these bistable cases. Note that different assumptions in the numerical models 
(for instance variable transport properties) help to extend the dipolar regime 
towards higher density contrasts \cite[(Duarte et al. 2013)]{Duarte13}.

Although it is difficult to directly relate the diagnostic parameters employed 
in numerical models to their observational counterparts, the separation into 
two dynamo branches seems to be relevant to the sample of active M dwarfs 
displayed in Fig.~\ref{fig:dipM}. In particular, the late M dwarfs (with 
$M_\star < 0.15\,\text{M}_\odot$) seem to operate in two different dynamo 
regimes: the first ones show a strong dipolar field, while others 
present a weaker multipolar magnetic field with a pronounced time-variability.

This analogy between numerical models and observations of active M dwarfs
could be further assessed by additional observations. Indeed, if the analogy 
holds, stars with a multipolar field are expected over a continuous range of 
Rossby number where dipole-dominated large-scale fields are also observed (i.e.  
$0.01 < Ro_\text{emp} < 0.1$).

\begin{acknowledgments}
TG and LD are supported by the Special Priority Program 1488 ``PlanetMag'' of 
the German Science Foundation.
\end{acknowledgments}


\begin{thebibliography}{}

\bibitem[{{Christensen} \& {Aubert} (2006)}]{Christensen06}
{Christensen}, U.~R. \& {Aubert}, J. 2006, \textit{Geophys. J. Int.}, 166, 97

\bibitem[{{Christensen} (2010)}]{Christensen10}
{Christensen}, U.~R. 2010, \textit{Space Sci. Rev.}, 152, 565

\bibitem[{{Donati} {et~al.} (2006)}]{Donati06}
{Donati}, J.-F., {Forveille}, T., {Cameron}, A.~C., {et~al.} 2006, 
\textit{Science},
  311, 633

\bibitem[{{Donati} {et~al.} (2008)}]{Donati08}
{Donati}, J.-F., {Morin}, J., {Petit}, P., {et~al.} 2008, \textit{MNRAS}, 390, 
545

\bibitem[{{{Duarte} et~al.} (2013)}]{Duarte13} {Duarte} L., {Gastine} 
T., {Wicht} J., 2013, \textit{Physics of the Earth and 
Planetary Interiors}, 222, 22


\bibitem[{{Gastine} \& {Wicht} (2012)}]{Gastine12}
{Gastine}, T. \& {Wicht}, J. 2012, \textit{Icarus}, 219, 428

\bibitem[{{Gastine} {et~al.} (2012)}]{Gastine12a}
{Gastine}, T., {Duarte}, L., \& {Wicht}, J. 2012, \textit{A\&A}, 546, A19

\bibitem[{{{Gastine} et~al.} (2013)}]{Gastine13}
{Gastine} T.,  {Morin} J.,  {Duarte} L.,  {Reiners} A.,  {Christensen} U.~R.,
   {Wicht} J.,  2013, \textit{A\&A}, 549, L5

\bibitem[{{{Jones} et~al.} (2011)}]{Jones11}
{Jones} C.~A.,  {Boronski} P.,  {Brun} A.~S.,  {et al.}  2011, \textit{Icarus}, 
216, 120

\bibitem[{{Kiraga} \& {Stepien}(2007)}]{Kiraga07} {Kiraga}, M. \& {Stepien}, K. 
2007, \textit{Acta Astronomica}, 57, 149

\bibitem[{{Morin} {et~al.} (2008a)}]{Morin08a}
{Morin}, J., {Donati}, J.-F., {Forveille}, T., {et~al.} 2008a,
  \textit{MNRAS}, 384, 77

\bibitem[{{Morin} {et~al.} (2008b)}]{midM}
{Morin}, J., {Donati}, J., {Petit}, P., {et~al.} 2008b,  
\textit{MNRAS}, 390, 567

\bibitem[{{Morin} {et~al.} (2010)}]{lateM}
{Morin}, J., {Donati}, J.-F., {Petit}, P., {et~al.} 2010, \textit{MNRAS}, 407, 
2269

\bibitem[{{Schrinner} {et~al.} (2012)}]{Schrinner12}
{Schrinner}, M., {Petitdemange}, L., \& {Dormy}, E. 2012, \textit{ApJ}, 752, 121

\bibitem[{{Semel} (1989)}]{Semel89} {Semel}, M. 1989, \textit{A\&A}, 225, 456

\bibitem[{{Simitev} \& {Busse} 	(2009)}]{Simitev09}
{Simitev}, R.~D. \& {Busse}, F.~H. 2009, \textit{Europhysics Letters}, 85, 19001

\bibitem[{{Wicht} (2002)}]{Wicht02}
{Wicht}, J. 2002, \textit{Physics of the Earth and Planetary Interiors}, 132, 
281

\bibitem[{{{Yadav} et~al.} (2013)}]{Yadav13a} {Yadav} R.~K.,  {Gastine} T.,   
 {Christensen} U.~R., 2013a, \textit{Icarus}, 225, 185

\bibitem[{{{Yadav} et~al.} (2013)}]{Yadav13b} {Yadav} R.~K.,  {Gastine} T.,  
{Christensen} U.~R., {Duarte} L.~D.~V., 2013b, \textit{ApJ}, 774, 6


\end{thebibliography}
\end{document}